\documentstyle[12pt]{article}
\advance \textheight by \topskip
\begin{document}

\begin{titlepage}
\hbox to \hsize{\hfil IHEP 95--146}
\hbox to \hsize{\hfil hep-ph/9512418}
\hbox to \hsize{\hfil April, 1996}
\vfill
\large \bf
\begin{center}
On asymmetry in inclusive pion production
\end{center}
\vskip 1cm
\normalsize
\begin{center}
{\bf S. M. Troshin and N. E. Tyurin}\\
{\small Institute for High Energy Physics,\\
  Protvino, Moscow Region, 142284 Russia}
\end{center}
\vskip 2.cm
\begin{abstract}
On the basis of the mechanism proposed  for one-spin
asymmetries in inclusive hadron production we specify an
$x$--dependence of asymmetries in inclusive
processes of pion production. The main role in
generation of this asymmetry belongs to the orbital angular momentum
of  quark-antiquark cloud in  internal structure of
constituent quarks. The $x$--dependence of asymmetries
in the charged pion production at large $x$ reflects
the corresponding dependence of constituent quark polarization
in the polarized proton. \\
PACS number(s): 13.85.Ni, 11.30.Qc, 12.39.Jh, 13.88.+e
\end{abstract}
\begin{center}
Revised version, to be published in Phys. Rev. D.
\end{center}
\vfill
\end{titlepage}

\section{Introduction} As it is widely known now, only part (less
than one third in fact) of the proton spin is due to quark spins
\cite{ellis,altar}.  These results can be interpreted in the
effective QCD approach ascribing a substantial part of hadron spin
to an orbital angular momentum of quark matter.  It is natural to
guess that this orbital angular momentum might be revealed in
asymmetries in hadron production.  In the recent paper \cite{asy} we
considered a possible origin of asymmetry in the pion production
under collision of a polarized proton beam with unpolarized proton
target and argued that the orbital angular momentum of partons inside
constituent quarks  leads to significant asymmetries in hadron
production with polarized beam.  This model has been successfully
applied to an OZI-violating process of $\varphi$--meson production
\cite{str}.

Various mechanisms were proposed recently as a source of the
significant asymmetries observed in pion production:  higher twist
effects \cite{efremt}, correlation of $k_\perp$ and spin in structure
\cite{sivs} and fragmentation \cite{cols,artru} functions, rotation
of valence quarks inside a hadron \cite{boros}.  Significant role in
part of above references also belongs to orbital angular momentum.
Recent reviews of theoretical and experimental aspects of
single--spin asymmetry studies have been given in \cite{trev,erev}.

In the model \cite{asy} the behavior of asymmetries in inclusive
 meson production was predicted to have a specific
$p_{\perp}$--dependence, in particular, vanishing asymmetry at
$p_{\perp}<\Lambda_\chi $, its increase in the region of
$p_{\perp}\simeq\Lambda_\chi $, and $p_{\perp}$--independent
asymmetry at $p_{\perp}>\Lambda_\chi $.  Parameter
$\Lambda_\chi\simeq 1$ GeV/c is determined by the scale of chiral
symmetry spontaneous breaking.  Such a behavior of asymmetry  follows
from the fact that the constituent quarks themselves have slow (if at
all) orbital motion and are in the $S$--state, but interactions with
$p_{\perp}>\Lambda_\chi $ resolve the internal structure of
constituent quark and feel the presence of internal orbital momenta
inside this constituent quark.

 We consider a nonperturbative hadron to be consisting of the
  constituent quarks located at the core of the hadron and
quark condensate surrounding this core.  Experimental and theoretical
arguments in favor of such a picture were given,  e.g. in
\cite{isl,trtu}.  We refer to effective QCD and use the NJL model
\cite{njl} as a basis. The  Lagrangian  in addition to the
$4$--fermion interaction of the original NJL model includes
$6$--fermion $U(1)_A$--breaking term.

Strong interaction radius of this quark is determined by its Compton
wavelength:  $r_Q=\xi/m_Q$, where the constant $\xi$ is universal for
different flavors.  Spin of constituent quark $J_{U}$  in this
 approach is determined by the  sum $J_{U}=1/2=J_{u_v}+J_{\{\bar q
 q\}}+\langle L_{\{\bar qq\}}\rangle$.  The value of the orbital
 momentum contribution into the spin of constituent quark can be
 estimated with account for the  experimental results from
 deep--inelastic scattering.  The important point  what the origin of
 this orbital angular momentum is. It was proposed \cite{asy} to
  consider an analogy with an  anisotropic generalization of the
 theory of superconductivity which seems to match well with the above
 picture for a constituent quark.

 In Ref. \cite{asy} we described how the orbital angular momentum,
i.e.  orbital motion of quark matter inside constituent quark, can be
 connected with the  asymmetries in inclusive production at moderate
and high transverse momenta and have given the predictions for
$p_{\perp}$--dependence of asymmetry.  In this note we specify
$x$--dependence of asymmetries. In section 2 we give a brief outline
of the model, section 3 is devoted to the description of an
$x$--dependence of asymmetry in inclusive pion production and in
section 4 we give short conclusion.

\section{Outline of the model} We  consider the hadron processes of
 the  type \[ h_1^\uparrow +h_2\rightarrow h_3 +X\] with polarized
 beam or target and $h_3$ being a charged pion.

 In the model constituent quarks  are supposed to scatter in a
quasi-independent way by some effective field which is being
 generated at the first stage of interaction under overlapping of
peripheral condensate clouds \cite{asy,trtu}.  Inclusive production
  of hadron $h_3$  results from recombination of the constituent
 quark (low $p_{\perp}$'s, soft interactions) or from the excitation
 of this constituent quark, its decay and subsequent fragmentation in
 the hadron $h_3$. The latter process is determined by the
 interactions at distances smaller than constituent quark radius and
is associated therefore with hard interactions (high $p_{\perp}$'s).
Thus, we adopt a two--component picture of hadron production which
 incorporates interactions at large and small distances.  We supposed
  that $\pi^+$--mesons are produced mostly by up flavors and
$\pi^-$--mesons --- by down flavors.

In the expression for asymmetry $A_N$ \cite{asy} \begin{equation}
A_N(s,\xi)=\frac{\int_0^\infty bdb I_-(s,b,\xi)/|1-iU(s,b)|^2}
{\int_0^\infty bdb I_+(s,b,\xi)/|1-iU(s,b)|^2} \label{xnn}
\end{equation} the function $U(s,b)$ is the generalized reaction
matrix \cite{log} and the functions $I_{\pm}$ can be expressed
through its multiparticle analogs \cite{asy}, $\xi$ denotes
the set of kinematical variables for the detected meson.
  In the model the
spin--independent part $I_+(s,b,\xi)$ gets contribution from the
processes at small (hard processes) as well as at large (soft
processes) distances, i.e.  $I_+(s,b,\xi)= I^h_+(s,b,\xi)+
I^s_+(s,b,\xi)$, while the spin--dependent part $I_-(s,b,\xi)$ gets
contribution from the interactions at short distances only
$I_-(s,b,\xi)=I^h_-(s,b,\xi)$.  The function $I_-^h(s,b,\xi)$ gets a
nonzero value due to  interference between the two helicity
amplitudes, which gain different phases due to internal motion of
partons inside the constituent quark.  The following relation between
the functions $I_-^h(s,b,\xi)$ and $I_+^h(s,b,\xi)$ has been proposed
assuming  the effect of internal motion of partons inside constituent
quark  leads to a shift in the produced meson transverse momentum:
\begin{equation} I_-^h(s,b,\xi)= \sin[{\cal{P}}_{\tilde Q}(x) \langle
L_{\{\bar q q\}}\rangle] I^h_+(s,b,\xi), \end{equation} where
${\cal{P}}_{\tilde Q}(x)$ is the polarization of the leading
constituent quark $\tilde Q$ (in the process  of the meson $h_3$
production) and $\langle L_{\{\bar q q\}}\rangle$ is the mean value
of internal angular momentum inside the constituent quark. This
relation allowed to get a parameter--free prediction for the
$p_{\perp}$--dependence of asymmetries in inclusive pion production
\cite{asy}.
 We have considered there
  ${\cal{P}}_{\tilde Q}$ being a constant. Here we assume
${\cal{P}}_{\tilde Q}$ being a function of $x$ and consider the
behavior of asymmetry in the beam fragmentation region
 (where $x\simeq x_F$).

The $x$--dependencies of the functions $I_+^s(s,b,\xi)$ and
$I_+^h(s,b,\xi)$ are determined by the distribution of constituent
quarks in a hadron and by the structure function of constituent quark
respectively:  \begin{equation} I_+^s(s,b,\xi)  \propto
\omega_{\tilde Q/h_1}(x) \Phi^s(s,b,p_{\perp})\quad \mbox{and} \quad
I_+^h(s,b,\xi)\propto \omega_{\tilde q/\tilde Q}(x)
\Phi^h(s,b,p_{\perp}). \label{is} \end{equation}
Taking into account the above
relations, we can represent inclusive cross--section in the unpolarized
case
$d\sigma/d\xi$ and asymmetry  $A_N$ in the
following forms:
\begin{equation}
\frac{d\sigma}{d\xi}=8\pi
[W_+^s (s,\xi)+W_+^h(s,\xi)],\label{cs}
 \end{equation}

 \begin{equation} A_N(s,x,p_{\perp})=\frac{ \sin[{\cal{P}}_{\tilde Q}(x)
\langle L_{\{\bar q q\}}\rangle] {W_+^h(s,\xi)}}{ {
[W_+^s (s,\xi)+W_+^h(s,\xi)]}},\label{an}
 \end{equation}
where the  functions $
W_+^{s,h}$ are determined by the interactions at large and small
distances:  \[ W_+^{s,h}(s,\xi)=\int_0^\infty
bdb{I_+^{s,h}(s,b,\xi)}/ {|1-iU(s,b)|^2}.  \]
\section{The $x$-dependence of asymmetry} The asymmetry in the model
  has a
significantly different $x$-dependence in the regions of transverse
 momenta $p_{\perp}\leq \Lambda_\chi$ and $p_{\perp}\geq
\Lambda_\chi$. Therefore we consider these two kinematical regions
separately. For that purpose it is useful to introduce the ratio \[
R(s,\xi)=\frac{W_+^h(s,\xi)}{W_+^s(s,\xi)}= \frac{\omega_{\tilde
q/\tilde Q}(x)} {\omega_{\tilde Q/h_1}(x)} r(s,p_{\perp}), \] where
the function $r(s,p_{\perp})$ in its turn is the $x$--independent
ratio \[ r(s,p_{\perp}) =\frac{\int_0^\infty bdb
\Phi^h(s,b,p_{\perp})/|1-iU(s,b)|^2} {\int_0^\infty bdb
\Phi^s(s,b,p_{\perp})/|1-iU(s,b)|^2}.  \] The expression for the
asymmetry $A_N(s,\xi)$ can be rewritten in the form \begin{equation}
 A_N(s,x,p_{\perp})= \sin[{\cal{P}}_{\tilde Q}(x) \langle L_{\{\bar q
q\}}\rangle] {R(s,x,p_{\perp})}/ {[1 +R(s,x,p_{\perp})]},
\end{equation} The function $R(s,x,p_{\perp})\gg 1$ at
$p_{\perp}>\Lambda_\chi$ since in this region dominate short distance
processes  and due to the similar reason $R(s,x,p_{\perp})\ll 1$ at
$p_{\perp}\leq\Lambda_\chi$.  Thus we have simple
$p_{\perp}$--independent expression for asymmetry at
$p_{\perp}>\Lambda_\chi$ \begin{equation} A_N(s,x,p_{\perp})\simeq
 \sin[{\cal{P}}_{\tilde Q}(x) \langle L_{\{\bar q q\}}\rangle]
\label{las} \end{equation} and a more complicated one for
$p_{\perp}\leq\Lambda_\chi$ \begin{equation}
 A_N(s,x,p_{\perp})\simeq\sin[ {\cal{P}}_{\tilde Q}(x) \langle
L_{\{\bar q q\}}\rangle] \frac{\omega_{\tilde q/\tilde Q}(x)}
{\omega_{\tilde Q/h_1}(x)} r(s,p_{\perp}). \label{sas} \end{equation}
 As it is clearly seen from Eq. (\ref{sas}) the asymmetry at
$p_{\perp}\leq\Lambda_\chi$ has a nontrivial $p_{\perp}$--dependence.
In this region asymmetry vanishes at small $p_{\perp}$ and is
suppressed also by the factor\\ $\omega_{\tilde q/\tilde
Q}(x)/\omega_{\tilde Q/h_1}(x)$ which can be considered as the ratio
of sea and valence quark distributions in hadron.  The
 $x$--dependence of asymmetry in this kinematical region strongly
 depends on particular parameterization of these distributions. We
 therefore will  consider the region of transverse momenta
 $p_{\perp}>\Lambda_\chi$ where the $x$--dependence of asymmetry has
 a simple form reflecting corresponding dependence of leading
 constituent quark polarization.  In \cite{asy} the two different
 cases were considered ${\cal{P}}_U=2/3$, ${\cal{P}}_D=-1/3$ and
 ${\cal{P}}_U= -{\cal{P}}_D=1$ .  The latest choice seems to be in
 better agreement with the experimental data. It is evident also that
 the experimental data \cite{E704} and Eq. (\ref{las}) point out
 increase of ${\cal{P}}_{\tilde Q}(x)$ with $x$. Therefore we take
  the above values of constituent quark polarization as the maximal
 ones and consider the simplest possible dependencies, e.g.  linear
 and quadratic ones:  \begin{equation} {\cal{P}}_{\tilde Q}(x)=
 {\cal{P}}_{\tilde Q}^{max}x\quad \mbox{and}\quad {\cal{P}}_{\tilde
 Q}(x)= {\cal{P}}_{\tilde Q}^{max}x^2, \end{equation} where
 ${\cal{P}}_U^{max}= -{\cal{P}}_D^{max}=1$.  The curves corresponding
 to the above parameterizations are presented in Figs. 1 and 2 by
 dashed lines. The  value of $\langle L_{\{\bar q q\}}\rangle\simeq
 1/3$ has been taken \cite{asy} on the basis of the analysis
 \cite{ellis} of the DIS experimental  data. As it is seen from Figs.
 1 and 2 the mean orbital angular momentum seems to be underestimated
 and experimental data prefer its greater value. Indeed, another
 analysis of DIS experiments has been carried out in \cite{vos} and
 the smaller value of the total spin carried by quarks has been
 obtained $\Delta\Sigma\simeq 0.2$. This value corresponds to
 $\langle L_{\{\bar q q\}}\rangle\simeq 0.4$.  Using the above value
 of angular orbital momentum we obtain a good agreement with the data
 in the case of linear dependence of constituent quark polarization
 and better but still not good agreement in the case of quadratic
 dependence of this polarization on $x$. The corresponding curves are
 presented in Figs. 1 and 2 by solid lines. It seems that the
 existing experimental data prefer linear dependence on $x$ of
 constituent quark polarization.  Note also that in this model the
 mirror symmetry for asymmetries in $\pi^+$ and $\pi_-$ production at
 $p_{\perp}>\Lambda_\chi$, i.e. $A_N^{\pi^+}=-A_N^{\pi^-}$ takes
 place.  Due to this relation asymmetry in $\pi^o$ production
 $A_N^{\pi^o}>0$ since from experimental data it is known that
$d\sigma^{\pi^+}/d\xi> d\sigma^{\pi^-}/d\xi$.

To perform the quantitative description of $x$--dependence of
 inclusive cross--sections of the pion production in the collisions
of unpolarized hadrons we should choose according to Eqs.
(\ref{is}), (\ref{cs}) the particular parameterizations of the
functions
$\omega_{\tilde Q/h_1}(x)$ and $\omega_{\tilde q/\tilde Q}(x)$.
As it was noted these functions could be associated with the valence
and sea quark distributions respectively. Noting that we  use the simplest
forms \cite{gunion} appropriate for fragmentation region:
\begin{equation}
\omega_{U/p}\sim (1-x)^3,\quad \omega_{D/p}\sim (1-x)^4
\end{equation}
and
\begin{equation}
\omega_{u/U}(x)\sim\omega_{d/D}(x)\sim (1-x)^5.
\end{equation}
Then for the cross-sections of inclusive $\pi ^+$- and $\pi ^-$-
production we have:
\begin{eqnarray}
x{\frac{d\sigma}{dx}}^{\pi^+} & = & C_V^{\pi^+}(1-x)^3+C_S^{\pi^+}(1-x)^5,
\label{pp}\\
x{\frac{d\sigma}{dx}}^{\pi^-} & = & C_V^{\pi^-}(1-x)^4+C_S^{\pi^-}(1-x)^5,
\label{pm}
\end{eqnarray}
where the factors $C_{V,S}^{\pi^\pm}$ are the constants at fixed energy.
Using the available experimental data at $p_L=400$ GeV/c \cite{nad}
we obtain a good agreement of Eqs. (\ref{pp}), (\ref{pm}) with experiment
at $0.2<x<0.7$ with the following values of the factors
$C_V^{\pi^+}=20$ mb,
$C_V^{\pi^-}=10$ mb,
$C_S^{\pi^+}=1$ mb,
$C_S^{\pi^-}=3$ mb.
The corresponding results are represented in Fig. 3.
\section{Conclusion} We have specified in this note the
$x$--dependence of asymmetry in inclusive pion production in the beam
fragmentation region in the framework of the approach formulated in
\cite{asy}. We assume that the polarization of constituent quark in
the polarized proton has a nontrivial $x$--dependence.  The
predictions for asymmetries given earlier in \cite{asy} should be
referred as to the fragmentation region.  The above considerations suggest
simple linear dependence of ${\cal{P}}_{\tilde Q}(x)$ at large $x$.
The main role in the generation of asymmetry belongs to the orbital
angular momentum of current quarks inside the constituent one.
Present considerations and the experimental data suggest that
$\langle L_{\{\bar q q\}}\rangle\simeq 0.4$.

We would like to conclude noting that the result of E143
 Collaboration on the measurements of $g_2(x)$ function suggests
small twist-3 matrix element \cite{g2} and the experimental result of
ALEPH Collaboration \cite{alph} reveals unexpectedly small
polarization of $\Lambda_b$  measured in $e^+e^-$--interactions in
the $Z^o$-decay which implies a strong depolarization mechanism
acting at the stage of fragmentation.  These interesting results
could have important impact on mechanisms of generation of spin
asymmetries in hadron production.  However, further experimental
studies are needed to clarify the spin puzzles observed in the
 measurements of asymmetries in hadronic processes.
\section*{Acknowledgements} We are thankful to D.  Kharzeev, W.-D.
Nowak, V. Petrov for stimulating discussions and to A. Vasiliev for
comments on E-704 experimental data.
  \newpage
\section*{Figure Captions} \bf Fig. 1 \rm \\ Asymmetries $A_N$ in
$\pi^+$ (positive values) and $\pi^-$(negative values) production in
$pp$--collisions at $p_L=200$ GeV/c. Curves correspond to linear
 dependence on $x$ of constituent quark polarization. Solid lines
correspond to $\langle L_{\{\bar q q\}}\rangle\simeq 0.4$ and dashed
 lines --- to $\langle L_{\{\bar q q\}}\rangle\simeq 0.33$.\\[2ex]
\bf Fig. 2 \rm \\ Asymmetries $A_N$ in $\pi^+$ (positive values) and
$\pi^-$(negative values) production in $pp$--collisions at $p_L=200$
GeV/c. Curves correspond to quadratic dependence on $x$ of
 constituent quark polarization. Solid lines correspond to $\langle
 L_{\{\bar q q\}}\rangle\simeq 0.4$ and dashed lines --- to $\langle
 L_{\{\bar q q\}}\rangle\simeq 0.33$.\\[2ex]
\bf Fig. 3 \rm \\ Inclusive cross--sections of $\pi^+$ (solid line) and
$\pi^-$ (dashed line) production in $pp$--collisions at $p_L=400$
GeV/c.

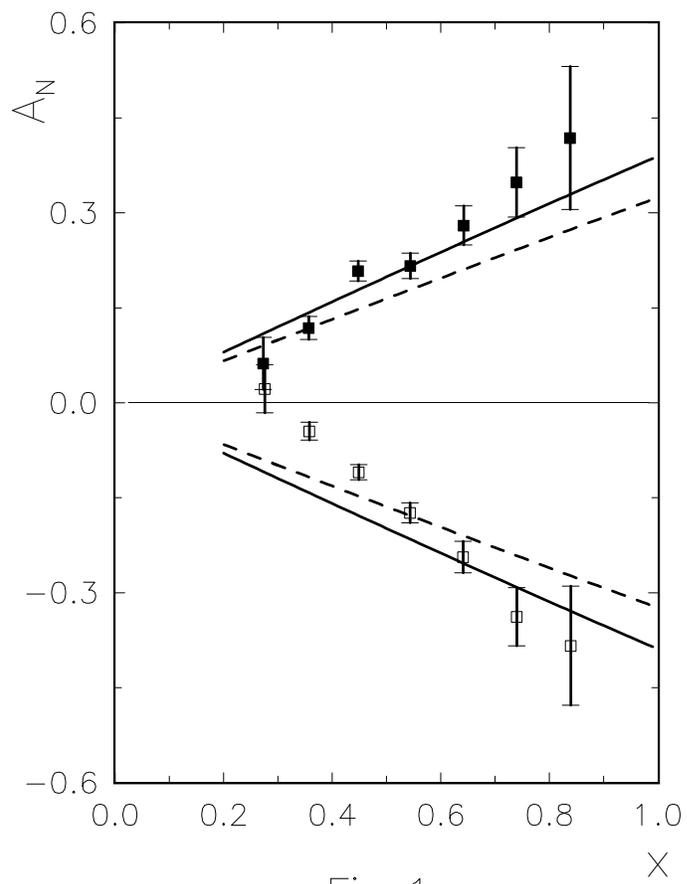
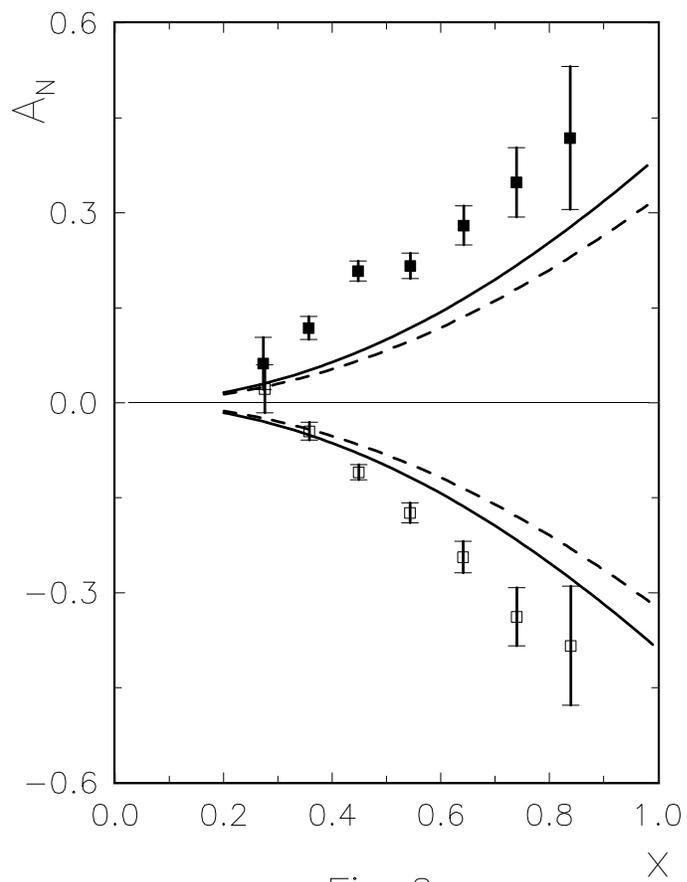
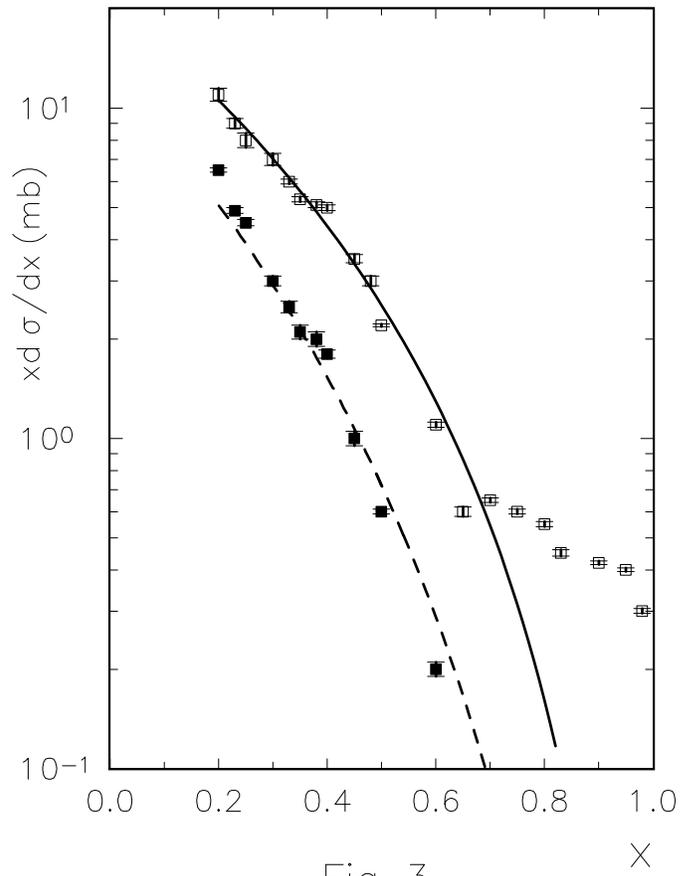
\end{document}